\documentclass[reprint,onecolumn,notitlepage,superscriptaddress,amsmath,amssymb,aps,prb,showpacs]{revtex4-1}
\usepackage[]{graphicx,subfigure}
\usepackage{dcolumn}
\usepackage{bm}
\usepackage{hyperref}
\usepackage{multirow}
\usepackage{tabularx}
\usepackage{siunitx}

\usepackage[utf8]{inputenc}

\begin{document}

\title{On uncertainties in the reconstruction of nanostructures in EUV scatterometry and grazing incidence small-angle X-ray scattering}

\author{A. Fern\'{a}ndez Herrero}
\email{analia.fernandez.herrero@ptb.de}
\affiliation{Physikalisch-Technische Bundesanstalt (PTB), Abbestr. 2-12, 10587 Berlin, Germany}
\author{V. Soltwisch}
\affiliation{Physikalisch-Technische Bundesanstalt (PTB), Abbestr. 2-12, 10587 Berlin, Germany}
\author{M. Pflüger}
\affiliation{Physikalisch-Technische Bundesanstalt (PTB), Abbestr. 2-12, 10587 Berlin, Germany}
\affiliation{Potsdam Institute for Climate Impact Research (PIK) e. V.,
Telegrafenberg A 31, 14473 Potsdam, Germany\\}
\author{J. Puls}
\affiliation{Physikalisch-Technische Bundesanstalt (PTB), Abbestr. 2-12, 10587 Berlin, Germany}
\author{F. Scholze}
\affiliation{Physikalisch-Technische Bundesanstalt (PTB), Abbestr. 2-12, 10587 Berlin, Germany}

\begin{abstract}
Increasing miniaturization and complexity of nanostructures require innovative metrology solutions with high throughput that can assess complex 3D structures in a non-destructive manner. 
EUV scatterometry is investigated for the characterization of nanostructured surfaces.
The reconstruction is based on a rigorous simulation using a Maxwell solver based on finite-elements and is statistically validated with a Markov-Chain Monte Carlo sampling method.
Here it is shown that this method is suitable for the dimensional characterization of the nanostructures and the investigations of oxide or contamination layers. In comparison to grazing-incidence small-angle X-ray scattering (GISAXS) EUV allows to probe smaller areas. The influence of the divergence on the diffracted intensities in EUV is much lower than in GISAXS, which also reduces the computational effort of the reconstruction.
\end{abstract}
\maketitle
\section{Introduction}

Measurement and characterization of nanofeatures correspond to more than~50~\% of the manufacturing process of the integrated electronic circuits~\cite{orji_metrology_2018}.
Metrology methods must address the structural complexity of the new nanostructures and provide accurate and efficient information.
In-line metrology methods that are commonly used  are critical dimension scanning electron microscopy (cd-SEM), critical dimension atomic force microscopy (cd-AFM) and optical scattering techniques, such as optical critical dimension (OCD)~\cite{bunday_hvm_2016, orji_metrology_2018}.
Atomic force microscopy (AFM) and scanning electron microscopy (SEM) are commonly used.
However, SEM cross-section analysis might involve the destruction of the target or a complex data analysis from top view images~\cite{Roman_height_2004,villarrubia_scanning_2015}.
Atomic force microscopy (AFM) is limited by the accessibility of the tip, with the manufacturing and characterization of this posing a challenge due to the small dimensions~\cite{Dai_2020_accurate}.
This data is also subjected to the deconvolution of the signal with the tip shape~\cite{dai_measurements_2013}.
As an alternative to those methods, photon scattering does not destruct the sample and delivers ensemble average information.
Optical scatterometry has been already implemented as in-line metrology for the control of each step in multiple patterning lithographic processes~\cite{Calaon_scatterometry_2018}.
However, when shrinking the dimensions, the attainable resolution is not sufficient~\cite{sunday_chapter_2017}.
Small angle X-ray scattering (SAXS) uses wavelengths shorter than the structure sizes and can be used for probing the inhomogeneities in the electron density within a sample system. By tilting the sample, periodic nanostructures can also be analyzed~\cite{hu_small_2004, wang_small_2007}. 
However, the investigation of the nanostructured surface can profit from the grazing-incidence illumination.
Grazing-incidence small angle X-ray scattering (GISAXS) has larger sensitivity to the surface.
The advantages of GISAXS are ensemble results in short measurement times also on thick, non-homogeneous substrates.
GISAXS has already been used for probing periodic nanostructures~\cite{tolan_x-ray_1995,metzger_nanometer_1997,jergel_structural_1999, mikulik_coplanar_2001, yan_intersection_2007, suh_characterization_2016,soltwisch_reconstructing_2017,pfluger_grazing-incidence_2017,fernandez_herrero_applicability_2019, pfluger_extracting_2020}.

However, the main disadvantage of GISAXS in comparison to previous scattering methods is the large footprint of the beam projected into the sample. This is usually larger than any investigated target~\cite{pfluger_grazing-incidence_2017}.
Larger angles of incidence might be used to reduce the elongation of the footprint. To counteract the loss in sensitivity to the surface, this should be accompanied by the illumination by longer wavelengths of the photon beam. Moreover, with the advent of lab EUV sources, this method present a real alternative to scanning methods~\cite{nguyen_coherent_2018, Tanksalvala_non_2021}. 
We report, on the dimensional reconstruction of a lamellar grating using EUV scattering and its comparison to a GISAXS reconstruction. 
A dimensional reconstruction of a lamellar grating previously done using GISAXS~\cite{soltwisch_reconstructing_2017} is here revisited, accounting for overseen sources of uncertainties~\cite{fernandez_herrero_applicability_2019}. 
The estimation of the parameter uncertainties is done by a Markov Chain Monte Carlo method~\cite{foreman-mackey_emcee:_2013}. The results from both methods are generally in good agreement. However, EUV scattering has several advantages over GISAXS. EUV allows to reduce the footprint on the targets. Also, computation times are reduced, as the divergence does not contribute to the diffracted intensities as much as in GISAXS. Finally, EUV allows the identification and characterisation of oxide or contamination layers without the need for extra measurements. 
 
\section{Experimental details}

The experiments were conducted at the PTB's soft X-ray beamline (SX700)~\cite{scholze_high-accuracy_2001} and the four-crystal monochromator (FCM) beamline~\cite{krumrey_high-accuracy_2001} at the electron storage ring BESSY II.
The SX700 beamline covers the energy range from 50 eV to 1800 eV and it is completely under UHV. The FCM covers a photon energy range from 1.75 keV to 10 keV.

\begin{figure*}[h]
   \begin{center}   
   \includegraphics[width=.7\linewidth]{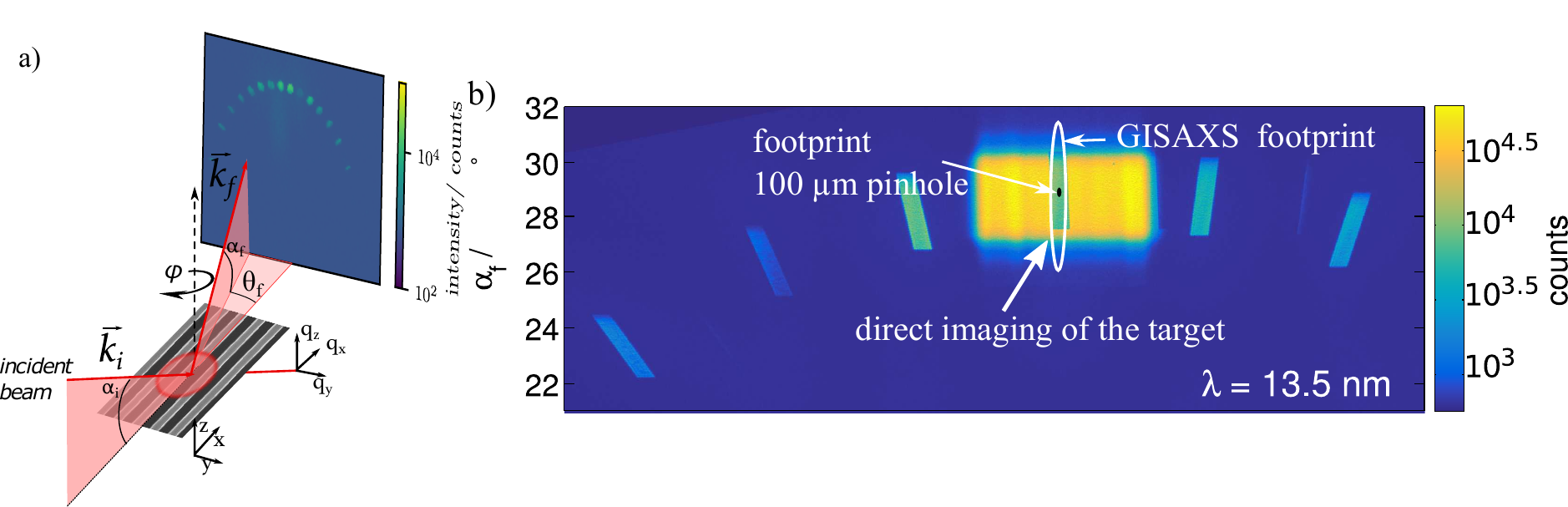}
   \caption{\label{fig:sketch}
\textbf{a)} Experimental set-up for EUV small angle scattering. \textbf{b)} CCD image of the grating diffraction rods obtained from a lamellar grating. The collimated incident beam allows a direct imaging of the sample area. Target areas can easily be identified by mapping the sample surface. The footprint between GISAXS and EUV is shown. By introducing a~100$\mu$m  pinhole in the beam path. The probed area is reduced to 0.1 mm horizontally by 0.2 mm vertically.}
\end{center}
\end{figure*} 

The SX700 beamline is designed for a beam with small divergence (regularly below~1~mrad) and minimal halo. For the investigations, the EUV angle resolved scatter set-up~\cite{fernandez_euvars_2018} is used to measure across a wide solid angle of the scattered light by placing the CCD sensor close to the sample. 
This experimental set-up is illustrated in Fig.~\ref{fig:sketch}. A monochromatic EUV beam with a wavevector $\vec{k}_i$ impinges on the sample surface at a grazing incidence angle $\alpha_i$. The elastically scattered wavevector $\vec{k}_f$ propagates with an exit angle $\alpha_f$ and an azimuth angle $\theta_f$. 
And the scattering vector is $\vec{q}=\vec{k}_f - \vec{k}_i$,

\begin{equation}
	\left(\begin{array}{c} q_x \\ q_y \\ q_z \end{array}\right)=k_0\left( \begin{array}{c} \cos(\theta_f)\cos(\alpha_f)-\cos(\alpha_i) \\
	 \sin(\theta_f)\cos(\alpha_f) \\
	 \sin(\alpha_f)+\sin(\alpha_i) \end{array} \right)\\	
\label{eq:qvector}
\end{equation}

The sample can be rotated around $\varphi$, with the grating lines parallel to the scattering plane for $\varphi=0$. A back-illuminated CCD camera is used. It has a 2048x2048 pixel area with a pixel size of 13.5$\mu$m.
The detector is placed close to the sample, at about $40$mm, allowing exit angles of $\sim 30^\circ$. 
This set-up has already been reported elsewhere~\cite{fernandez_euvars_2018}.
By using a collimated beam large areas can be investigated (see Fig.~\ref{fig:sketch}b)). This can be used for the detection of large inhomogenities on the samples~\cite{fernandez_euvars_2018} and to navigate the sample. The footprint can be reduced by using a set of pinholes. Here, a 100 $\mu$m pinhole was used, which allows the investigation of smaller targets. Figure~\ref{fig:sketch} b) shows the footprint comparison between GISAXS and EUV.
Compared to GISAXS, this set-up allows the heavy reduction of the footprint of the beam onto the targets. 
In order to increase the amount of information about the nanostructures, the reciprocal space was mapped. For this, the energy of the incoming photon beam was tuned from 200 eV to 220 eV in steps of 1 eV. 

The conical mounting was also used in the FCM beamline, where the acquisition of GISAXS was done. A beam-defining pinhole of about 500 $\mu$m was used at a distance of about 1.5 m to the sample. Together with a scatter guard of 1000 $\mu$m close to the sample, the beam spot size was about 0.5 mm $x$ 0.5 mm at the sample position. The beam had a horizontal divergence of 0.01$^\circ$ and a vertical divergence of 0.006 $^\circ$.
The detector is an in-vacuum PILATUS 1M detector~\cite{wernecke_characterization_2014} with a pixel size of $(172$ x $172)$ $\mu$m$^2$, placed at a distance of about 3.5 m. The incidence angle is approximately $\alpha_i = 0.85^\circ$. In order to obtain more information on the structures, several parts of the reciprocal space were mapped by varying the photon energy.
The set-up looks similar to the one show in Fig.~\ref{fig:sketch} a) but with smaller incidence angle $\alpha_i$, which leads to the elongation of the footprint.

The sample consists of a state-of-the-art Si-lamellar grating produced by e-beam lithography. It has a pitch of 150 nm, a nominal height of 120 nm and a nominal line width of 65 nm. This structure was previously reconstructed using GISAXS without accounting for the vertical divergence of the beam, which is actually the leading error contribution~\cite{fernandez_herrero_applicability_2019}. The reconstruction using GISAXS is revisited here using a more versatile method to account for the uncertainties and thereby, a more reliable confidence interval estimation.

\section{Characterization of the line shape}

The reconstruction of nanostructures from the diffracted intensities corresponds to an optimization problem based on the forward calculation of the experimental realization (including the model of the line).
The computation of the forward model can be done by different methods~\cite{popov_gratings:_2014}. Distorted-wave Born approximation has been broadly used for the characterization of the nanostructures~\cite{Babonneau:hx5104,Lazzari:vi0158,renaud_probing_2009,rueda_grazing-incidence_2012,rauscher_small-angle_1995,jiang_waveguide-enhanced_2011-1,hofmann_grazing_2009,suh_characterization_2016,meier_situ_2012}. Nevertheless, 
they are not as reliable as rigorous methods when the exact intensity distribution is pursued~\cite{Pflueger_2020_Using}.
Rigorous methods, such as Maxwell solvers based on finite elements, have already been used for the characterization of periodic nanostructures~\cite{soltwisch_reconstructing_2017,fernandez_herrero_applicability_2019, pfluger_extracting_2020}.
The faithful reconstruction of a nanostructure relies on a good describing model of the experimental set-up.
It has been reported before that the divergence of the light has a big impact on the diffracted intensities from GISAXS and thus, in the reconstruction of the nanostructures~\cite{soltwisch_reconstructing_2017, fernandez_herrero_applicability_2019}.
As well, the reconstruction of an error model in the optimization process allows the derivation of unbiased error budgets~\cite{fernandez_herrero_applicability_2019}.

For the computation of the diffracted intensities a Maxwell solver based on the finite element method (FEM) is used. It allows the rigorous computation of the near field distribution of any arbitrary shape. By a post-process based on a Fourier transformation the intensity of the diffraction orders is obtained. Here, the commercial software~\textit{JCMsuite} is used~\cite{burger_jcmsuite:_2008}.
To sample the posterior distribution a Markov Chain Monte Carlo method is used.
The posterior probability of the parameters depends on previous knowledge on the distribution of the parameters, which is the prior function, and on the likelihood function of the set of parameters. 

The likelihood is given by, 
\begin{equation}
    \vec{L} (\sigma,\vec{e})= \prod_{E,m} {[2\pi\sigma^2(m,E)]^{-1/2}}  \exp(-\chi^2(\vec{e},m,E)/2),
    \label{eq:likelihood}
\end{equation}

and $ \chi^2(\vec{e},m,E)$ corresponds to.

\begin{equation}
    \chi^2(\vec{e},m,E)= \frac{\left[ I_{calc}\left(\vec{e},m,E\right)-I_{exp}\left(m,E\right)\right]^2}{\sigma^2(m,E)}.
\label{eq:chi}
\end{equation}
where $m$ is order of diffraction, E is the energy and $\vec{e}$ are the variable parameters that are reconstructed. Usually the incoming incidence angle and the azimuthal rotation are included in the reconstruction together with the paramters defining the line shape. The distribution of the photon energy is also considered.

The calculated intensity $I_{calc}$ includes a Debye-Waller factor to account for the effect of the roughness on the diffraction orders~\cite{ kato_effect_2010, fernandez_herrero_applicability_2019}, $I_{calc} = I_{FEM} \exp(-q_y ^2 \xi^2 )$.
And $I_{FEM}$ is the computed intensity using the Maxwell solver.
The standard deviation $\sigma (m,E)$ is composed by the experimental and computational errors. 
It has been reported that the reliable derivation of confidence intervals rely on a good determination of the uncertainties contributing to the methods. 
In order to account for possible unknown contribution, an error model must be used~\cite{fernandez_herrero_applicability_2019}. For virtual scattering experiments, an Gaussian error model has been identified by Heidenreich et al.~\cite{Heidenreich_bayesian_2015}. The cases of EUV and GISAXS are discussed separately.


\subsection{GISAXS}

The description of the divergence of the light is indispensable to obtain an unequivocal solution in the reconstruction of nanostructures using GISAXS~\cite{fernandez_herrero_applicability_2019}.
Including the divergence in the computation increases the computational times by more than twenty times compared to a computation where the divergence can be disregarded. 
Different angles must be computed separately and convoluted with the profile of the beam for each model computation.
Even including the divergence in the reconstruction, approximations must be done.
Theses approximations are leading the contribution to the total uncertainty of the method~\cite{soltwisch_correlated_2016,fernandez_herrero_applicability_2019}. The impact of the divergence on the intensity for a comparable GISAXS experiment has been reported elsewhere~\cite{fernandez_herrero_applicability_2019}.
Depending on whether the orders lie on the reciprocal space (qz,qy,qx), the impact of the divergence is also different.
So that, when different energies are considered, an error must be fitted for each of them,
\begin{equation}
    \sigma^2(m,E) \approx \sigma^2_{model}(m,E) + \sigma_{exp}(m,E) = [a(E)I(m,E)]^2 + b^2(E) + \sigma_N(m,E) + \sigma_{hom}(E).
    \label{eq:error_gisaxs}
\end{equation}

The experimental error is given by two known errors. One is the detector inhomogenity ($\sigma_{hom}$)~\cite{wernecke_characterization_2014}, which is about $2 \%$.
Additionally, an uncertainty following the Poisson statistical distribution is considered $\sigma_N(m, E)$, where m is the order of diffraction, and E is the energy.

The factor $b$ can be usually considered negligible and be omitted from the reconstruction~\cite{fernandez_herrero_applicability_2019, pfluger_extracting_2020}. 
However, here it was reconstructed to allow the identification of overseen contributions.
The reconstruction was done by using three different photon energies. No more energies were added to not further increases the computational time. 
Although the incoming intensity is known, the efficiency of the diffraction orders is not completely known. The footprint of the beam is larger than the target.
Therefore a scale must be included in the reconstruction. 
\begin{figure}[h]
  \hspace*{-2cm}
   \begin{center}   
   \includegraphics[width=1.\linewidth]{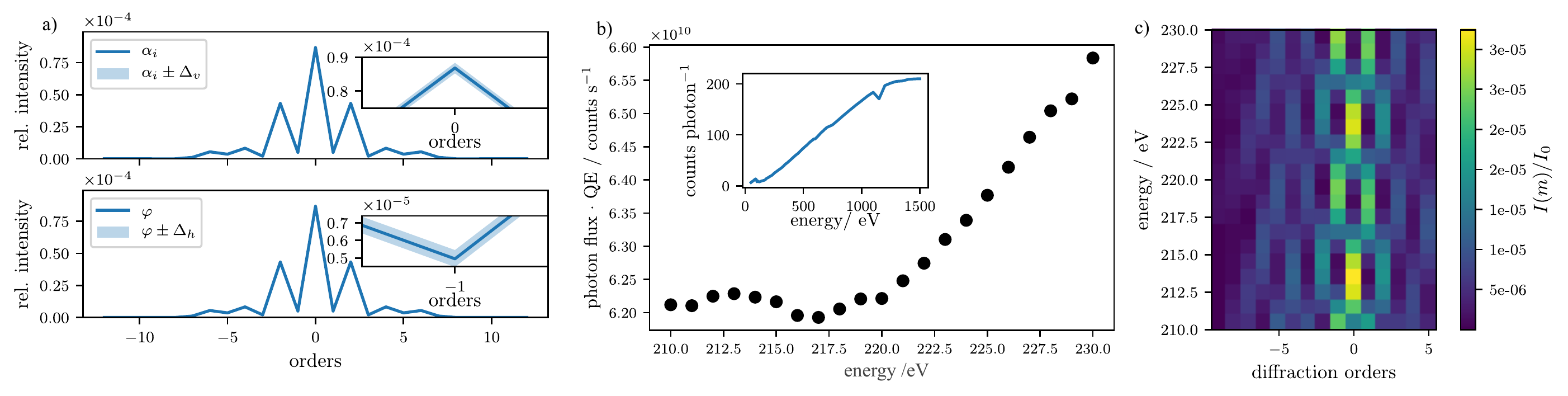}
   \caption{\label{fig:divergence}
\textbf{a)} Effect of the divergence in the intensity of the diffraction orders.  \textbf{b)} Conversion of the number of counts per photon per second. The inset shows the quantum efficiency of the CCD (counts per photon) for the whole energy range of the beamline. \textbf{c)} Diffraction efficiency of the diffraction orders as a function of the incident energy. }
\end{center}
\end{figure} 

\subsection{EUV scatterometry}

For GISAXS the effect of the divergence is leading the uncertainty contribution but also the computational time.
Thereby, the impact of the divergence on the diffracted intensities was studied previous its inclusion in the computation. 
Figure~\ref{fig:divergence} a) show the divergence of the light for this experiment in the EUV and with an incident angle of $\sim 30^\circ$.
The divergence effect can be left out of the reconstruction process, which leads to a reduction in the computational time of each structure and experimental set-up. 
The error model in this case is,
\begin{equation}
    \sigma^2(m,E) \approx \sigma^2_{model}(m) + \sigma_N(m,E) = [a\cdot I(m)]^2 + b^2 + \sigma_N(m,E).
    \label{eq:error_euv}
\end{equation}

This reduces the number of the parameters considerably when including more energies in the reconstruction. 
The reconstructed Gaussian error model includes the contribution of the divergence, as exposed before, but also of the numerical precision.
This latter error is due to the assumption that must be made in the computation in order to have a solution in reasonable times.
Although differences in this error are expected~\cite{soltwisch_correlated_2016} when varying the energy, the mapped range is only of 20 eV. And therefore, just one error might be sufficient. 

The CCD was calibrated for the extraction of absolute intensities, see inset Fig.~\ref{fig:divergence} b). 
Although the direct beam cannot be measured directly with the CCD, the incoming intensity was measured before entering the chamber (before the pinhole). The measured signal at the detector during the experiments is proportional to the photon flux. Therefore, one scale factor for all the energies must be also reconstructed. Figure~\ref{fig:divergence} b) shows the conversion rate from counts at the detector to incoming photons.
The extracted diffraction efficiencies are shown in Fig.~\ref{fig:divergence} c).

The line model considered in the reconstruction was also varied. The lamellar grating was produced by plasma etching, which is usually producing an oxide layer on the top of the structure. GISAXS is not sensitive to this small oxide layer on the top but in EUV scatterometry, it must be considered in the reconstruction. 
Because oxide silicon usually has lower densities than bulk SiO$_2$, a weighting factor to the density of the oxide layer is also reconstructed.
However, the optical constants are not reconstructed and tabulated data is used and scaled with the density.

\section{Confidence intervals}

\begin{figure*}
\centering
    \includegraphics{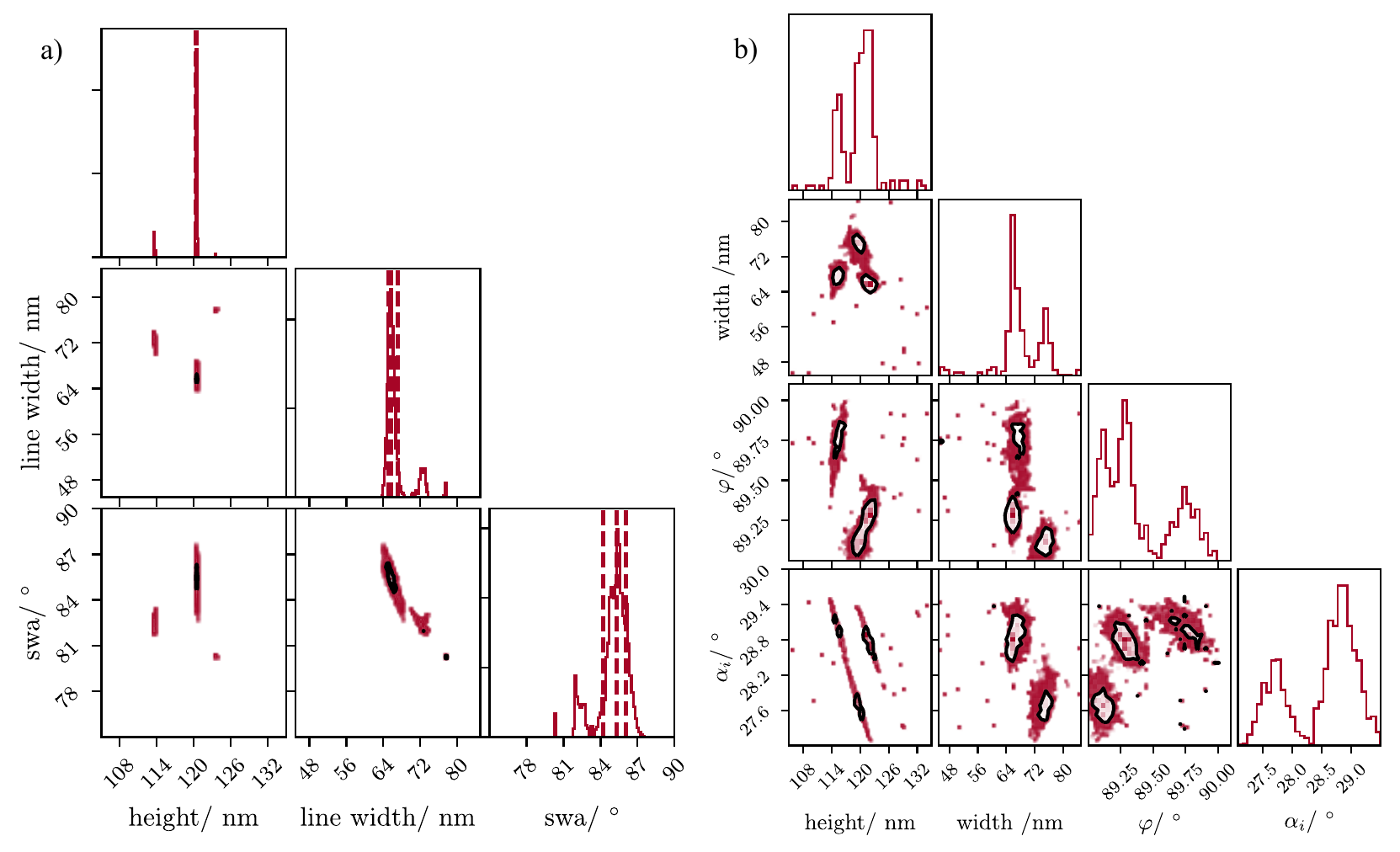}
    \caption{Corner plots showing low-dimensional projections of the posterior distribution, where the black contours show~68$\%$ of the mass of the projected posterior distribution. Please note that for 2D distributions, 1$\sigma$ corresponds to 39$\%$.  
    The posterior distribution of the main geometrical parameters is shown for GISAXS (\textbf{a)}). Almost all the solutions are found under one modality of the posterior.
    For EUV (\textbf{b)}) the low-dimensional projection of the posterior is showing the dependence of the results on the incident angles.}
    \label{fig:posterior}
\end{figure*}
The reconstruction was done by MCMC, using the emcee python package~\cite{foreman-mackey_emcee:_2013}. The analysis of the posterior distribution allows the investigation of the confidence intervals of the parameters. The search space was the same for both methods, except for the incident angles ($\varphi$, $\alpha_i$). The search space for each parameter can be seen in Table~\ref{tab:comparison-euv_gisaxs} (limits column).
The posterior distributions of both methods have multiple modalities. Although more experimental data points were included in the EUV reconstruction, an unequivocal solution in the whole searched space is not found. The relatively large search space for the reconstruction of the incidence angles can be responsible for this behaviour, see Figure~\ref{fig:posterior} b).
In GISAXS the angles are usually also reconstructed because diffracted intensities are very sensitive to any variation on the set-up~\cite{soltwisch_reconstructing_2017}. But the search space is restricted to $\pm 0. 02^\circ$ of the measured angle and does not lead to a multi-modal posterior.
However, in the EUV set-up~\cite{fernandez_euvars_2018} was not possible to know these angles and they were reconstructed allowing comparably large search spaces.
Actually the relative unknown incident angles have a strong influence on the observed modalities.
The multiple modalities are observed with a certain correlation between the angles and the parameters.
Here the width and height are shown.

For GISAXS the posterior distributions were also analysed. A slight cross-correlation is observed for the line width and sidewall angle, see Fig.~\ref{fig:posterior} a).
The line width is defined at the middle of the height, which can be responsible for this observed behaviour for a certain measurement set-up. Depending on the measurement geometry, the sensitivity of the parameters to the method might also change. 
Nevertheless, if the method were insensitive to one parameter, there would not be a defined solution of the posterior, which is here not the case.

The confidence intervals defined at 68 $\%$ of the mass for the GISAXS reconstruction are given in Table~\ref{tab:comparison-euv_gisaxs} (first GISAXS column). 
For a better comparison between the two results an area from the search space is selected. The nominal value of the height, 120 nm, is chosen. In this case, just one solution is found and we can speak about 1$\sigma$ of the uncertainty.

\section{Comparison between EUV scatterometry and GISAXS}

\begin{figure}
\centering
   \includegraphics[width=1.\linewidth]{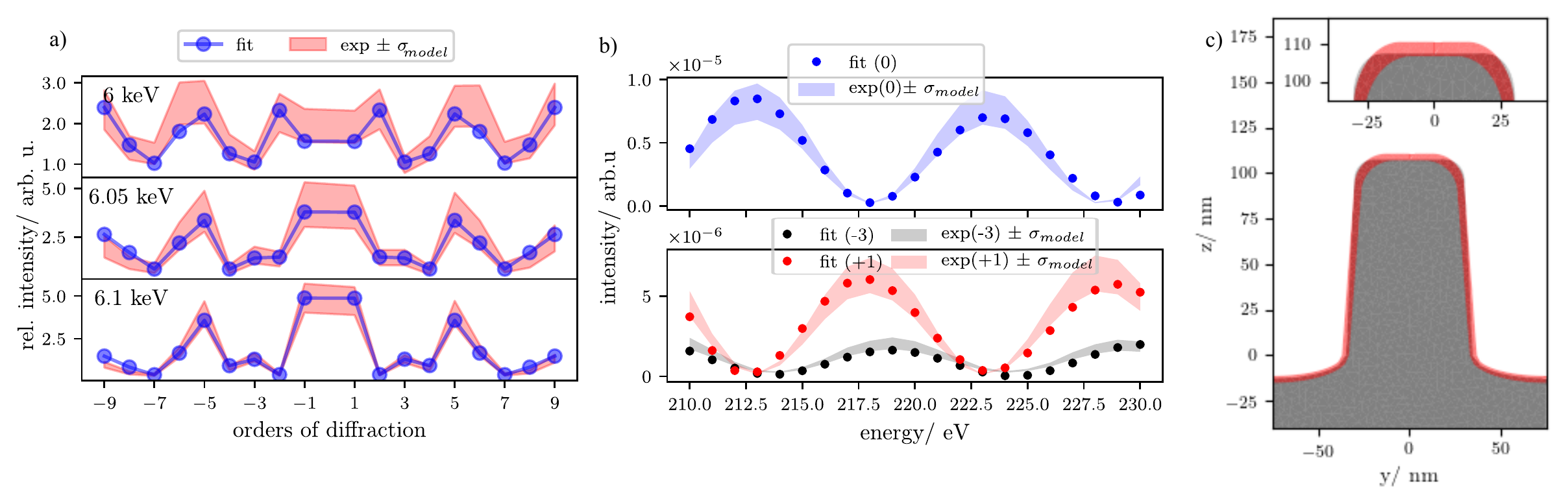}
   \caption{
\textbf{a), b)} Comparison between the measuremnt data and the best fit. The shaded area indicates the total error of the method (for visualization was applied to the experimental data), using GISAXS (\textbf{a)}) and using EUV scattering (\textbf{b)}). In this latter case, just the modulation of a few diffraction orders is shown.  \textbf{c)} The gray area corresponds to the reconstructed line shape using GISAXS, on the top, a SiO$_2$ layer (semi-transparent red) corresponds to the line profile and oxide layer reconstructed using EUV scatterometry. A slightly difference on the height and on the rounding is observed.}
\label{fig:fit}
\end{figure}

Figure~\ref{fig:fit} shows the comparison between the best fit and the measured data for each case: GISAXS (a)) and  EUV (b)).
The structure is reconstructed using an analogous method.
However, the layout of EUV was changed to allow the characterization of an oxide layer on the top of the structure.
The comparison between both layouts is shown in Fig.~\ref{fig:fit} c). Only a slightly different on the height is appreciated. 

For the comparison of both methods the search space is constrained around the nominal value of the height, that is $height  \in [117,123]$. The comparison of the values is given in Table~\ref{tab:comparison-euv_gisaxs}. 
Both methods are in agreement and deliver analogous result for the geometry of the structure.
The height is slightly different but the only parameter that is not within the uncertainty is the amplitude of the roughness that was reconstructed using a Debye-Waller factor.
The samples were measured with a time lapse of a few years, which could have cause the addition of a small contamination layer that is here not considered and that could have damped more the intensity of the diffraction orders. 
Another explanation is the difference between the two methods, each of them in an energy range. While in GISAXS the angle used is very small and the investigated surface is very large, in EUV is the opposite. Although EUV still delivers ensemble results, the meaurement is more local than GISAXS. It is also worth thinking whether the reconstructed roughness with each method corresponds to the same roughness. Although we have comparable $q_y$ ranges in the experiment, the $q_x$ component is completely different. 
Nevertheless, in the EUV energy range possible errors of the optical constants when assuming tabulated data cannot be completely ruled out.

The total error contributing to each method is also comparable between GISAXS and EUV.
In GISAXS no offset $b$ of the error is observed, which can be explained by the hybrid photon-counting detector.
Nevertheless, the uncertainty on the angles of the incoming beam in EUV must be addressed, to avoid the multiple modalities of the solution.

\begin{table}
    \centering
    \begin{minipage}{\linewidth}
    \centering
    \begin{tabular}{ l | c | c || c | c  }
    \hline
             & \multicolumn{2}{c||}{value $\pm$ confidence interval} & \multicolumn{2}{c} {\begin{tabular}{@{}c@{}} value $\pm$ 1 $\sigma$ \\ restricting $l_h$ = [117,123] \end{tabular}} \\ \hline
        parameter  & GISAXS & limits & GISAXS & EUV \\ \hline\hline
        height/ nm & $120. 49 \substack{ +0.07 \\ -0.12}$ & [105,135] &  120.49 $\pm$ 0.07 &   121.7 $\pm$ 0.8 \\
        height oxide / nm & - & - &  - & 3.89 $\pm$ 0.05  \\
        SiO$_{2}$ density & - & - & - & 0.78 $\pm$ 0.02  \\
        line width/ nm & $65.7 \substack{+1.4 \\ -0.7}$ & [45,85] &  65.7 $\pm$ 0.7 & 65.8 $\pm$ 0.8  \\
        swa/ $^{\circ}$ & $85.31 \substack{+0.8 \\ -1.12}$  &[75,90] & 85.41 $\pm$ 0.7 & 86.0 $\pm$ 0.5  \\
        top rounding/ nm & $10.3 \substack{+1.4 \\ -0.9}$ & [1,20] & 10.15 $\pm$ 0.8 & 17.1 $\pm$ 1.1 \\
        bottom etch/nm & $11.1 \substack{+0.9 \\ -0.9}$ & [1,15] & 11.2 $\pm$ 0.7 & 11.7 $\pm$ 0.3  \\
        $\xi$/ nm & $0.98 \substack{+0.3 \\ -0.3}$ & [0,7] & 1.0 $\pm$ 0.3 & 3.22 $\pm$ 0.13 \\ \hline
        a\footnote{the reconstructed error for GISAXS is given at 6 keV, plus the uncertainty of the detector. So that the total error of the method can be compared. In EUV this possible error was also reconstructed} / \%  & 15 $\pm$ 3 & [0,30] &  12 $\pm$ 3  & 15 $\pm$3 \\
        b\footnote{the obtained value for GISAXS correspond to less than one count per pixel}/ counts & - & - & - &  857 $\pm$ 97\\ \hline
        
    \end{tabular}
    \end{minipage}
    \caption{Values of the parameters defining the line profile obtained by the reconstruction done using MCMC from GISAXS data and EUV scatterometry. For GISAXS is shown the value and the confidence interval when considering 68 \% of the mass. For comparison (two columns on the right), an area around the nominal value is chosen.
    The confidence intervals are obtained from the posterior distributions, considering 1$\sigma$.
    SiO$_{2}$ density corresponds to a weighting factor that is fitted to the density, which is considered to be the bulk density of 2.2 g/cm$^{-3}$.}
    \label{tab:comparison-euv_gisaxs}
\end{table}

\section{Conclusions}
The reconstruction of a lamellar grating was done using GISAXS and EUV scatterometry. 
In comparison to GISAXS, EUV allows the investigation of smaller areas, which is of high interest for the investigation of small targets. 
The same reconstruction procedure has been applied to each of the data sets. 
Due to the high sensitivity of GISAXS to the photon beam divergence, this must be included in the reconstruction. However, for EUV this is not anymore an issue, which saves computation times for each set-up. 
Moreover, EUV allows the identification of small layers on the top of the structures.
Both methods show an overall good agreement between the reconstructed parameter values. Only the roughness amplitude which was fitted by including a Debye-Waller factor in the reconstruction delivers different values. 
This is subject to further investigations and might be due to a contamination layer, that has grown over time or to sensitivity to different roughness contributions of the respective methods. The smaller footprint and higher experimental robustness with regard to parameters such as photon beam divergence make EUV scattering a promising method for the characterization of nanostructures.

\label{sec:refs}

\bibliographystyle{unsrt}
\bibliography{literature, gratingseverywhere}

\end{document}